\documentclass[11pt,a4paper]{article}
\usepackage{jheppub}

\begin{document}

\thispagestyle{empty}

\hfill ULB-TH/11-05 \\

\title{Black Hole Thermodynamics and Massive Gravity}

\author[a]{Fabio Capela}
\author[a]{and Peter G. Tinyakov}
\affiliation[a]{Service de Physique Th\'{e}orique, 
Universit\'{e} Libre de Bruxelles (ULB),\\CP225 Boulevard du 
Triomphe, B-1050 Bruxelles, Belgium}

\emailAdd{fregocap@ulb.ac.be}
\emailAdd{petr.tiniakov@ulb.ac.be}

\abstract{We consider the generalized laws of thermodynamics in massive
gravity. Making use of explicit black hole solutions, we devise black
hole merger processes in which i) total entropy of the system
decreases ii) the zero-temperature extremal black hole is
created. Thus, both second and third laws of thermodynamics are
violated. In both cases, the violation can be traced back to the
presence of negative-mass black holes, which, in turn, is related to
the violation of the null energy condition. The violation of the third
law of thermodynamics implies, in particular, that a naked singularity
may be created as a result of the evolution of a singularity-free
state. This may signal a problem in the model, unless the creation of
the negative-mass black holes from positive-mass states can be
forbidden dynamically or the naked singularity may somehow be resolved
in a full quantum theory.}


\arxivnumber{1102.0479}

\maketitle

\section{Introduction}

There has been recently an increasing interest in infrared
modifications of gravity which could, in principle, explain the
accelerated expansion of the Universe without introducing a "dark
energy" component. However, before these models are taken seriously,
their consistency has to be checked both from theoretical point of
view and by comparison with the experimental data. One class of such
models involves spontaneous breaking of Lorentz symmetry by the
space-time dependent condensates of scalar fields coupled to gravity
through derivative couplings
\cite{ArkaniHamed:2002sp,ArkaniHamed:2003uy,Dubovsky:2004sg}. It has
been shown that such models may have a non-pathological behavior in
the perturbative regime and may exhibit infrared modifications of the
gravitational interactions. The models of this class are naturally
called "massive gravity" since the graviton generically acquires a
non-zero mass due to interactions with the scalar fields
\cite{Dubovsky:2004sg,Rubakov:2004eb}.  An important feature of these
models is that they are formulated in a non-perturbative way which
makes it possible to study non-linear solutions such as those
describing modified cosmological evolution \cite{Dubovsky:2005dw} or
black holes \cite{Dubovsky:2007zi,key-5} (for a review and further
references see, e.g., Ref.~\cite{Rubakov:2008nh}).

In the conventional General Relativity (GR), the existence of
solutions with horizons (e.g., black holes) raises questions of
consistency of a more general kind, namely, the consistency with the
general laws of thermodynamics. It has been argued
\cite{key-0.1,key-0.2,key-0.3} that in GR such consistency can be
obtained by assigning black holes a certain temperature and entropy. The
temperature of a black hole characterizes its Hawking radiation, so
that the black hole would be in a thermal equilibrium with the thermal
bath at this temperature. The entropy of a black hole was argued to be
proportional to its horizon area. In this way, the net entropy of the
black hole and the outer region never decreases, generalizing the
second law of thermodynamics to processes that include black holes.
These thermodynamical properties of black holes in GR are believed to
be connected to fundamental principles of quantum physics, such as
unitarity.  The validity of the thermodynamical description of black
holes should then give us some insight into quantum aspects of
gravity. For example, in string theory for a certain class of extremal
black holes the Bekenstein entropy has been reproduced by counting the
microscopic states of the compact objects \cite{key-25,key-26}.

A question that arises naturally is how the thermodynamic properties
of the black holes are changed in the modified gravity models, in
particular, in massive gravity, and whether these changes preserve
consistency with the thermodynamic laws. In the context of the ghost
condensate models \cite{ArkaniHamed:2003uy} this question has been
raised in Refs.~\cite{key-2,Feldstein:2009qy}. In Ref.~\cite{key-2} a
gedanken experiment involving a black hole has been proposed which
allows the transfer of heat from a cold body to a hot one, thus
violating the second law of thermodynamics (for the discussion of the
subtleties of the arguments see refs.\cite{key-3,key-4}). In
Ref.~\cite{Feldstein:2009qy} the violation of the second law of
thermodynamics was related to the presence of negative energy states.
On the contrary, in the context of the TeVeS models
\cite{Bekenstein:2004ne} it was conjectured \cite{key-7} that the
second law of thermodynamics holds if the effective graviton radiation
temperature and the Hawking radiation temperature are equal.

The aim of this paper is to check the validity of the laws of
thermodynamics in massive gravity by making use of the exact black
hole solutions of Ref.\cite{key-5} (solutions of this type were first
found in bi-metric models in
Ref.~\cite{Berezhiani:2008nr}).  Unlike conventional black holes,
these solutions depend on two parameters: the mass $M$ and the
``scalar charge'' $Q$.  We will concentrate on solutions which can be
interpreted as (modified) black holes, i.e., have a finite ADM mass
(and, therefore, produce the Newtonian $1/r$ potential at large
distances) and have an event horizon. We will first find the
expressions for the temperature and entropy of the modified black
holes. This will be done by calculating the surface gravity and by
making use of the Wald's representation of entropy as a Noether
charge. Having obtained the explicit expressions for the entropy and
temperature in terms of the black hole mass and scalar charge, we then
consider the process of merging of two black holes. As we will argue,
in the limit when one of the black holes is much smaller than the
other, the mass and the scalar charge of the resulting black hole is
the sum of masses and charges of the constituents. Thus, we will be
able to compare the entropies of the initial and final states and
check the second law of thermodynamics. We will see that when the
negative mass states are present, one can {\em decrease} the total
entropy during the merger. Moreover, one can create a zero-temperature
black hole --- an extremal state which, by an infinitesimal change of
initial parameters, can be converted into a state with naked
singularity. Thus, the second and third laws of thermodynamics can be
violated.

The paper is organized as follows. In Sect.~2 we review the static
spherically symmetric solutions of massive gravity. In Sect.~3 we
first compute the temperature and entropy of modified black holes in
massive gravity. We then verify the second and the third laws of
thermodynamics and show that both can be violated. Sect.~4 contains
the summary of the results and their discussion.

\section{Static Spherically Symmetric Solutions in Massive Gravity}

The massive gravity model which is used in this paper is described by
the following action \cite{Dubovsky:2004sg}:
\begin{equation}
  \mathcal{S}=\int dx^{4}\sqrt{-g}\left[-M_{pl}^{2}\mathcal{R}
  +\Lambda^{4}\mathcal{F}\right] 
\label{eq:5}
\end{equation}
where the first term is the usual Einstein-Hilbert action and the
second one is a certain function (to be specified below) of the
space-time derivatives of the four scalar fields $\phi^{0}$,
$\phi^{i}$ minimally coupled to gravity. This model should be viewed
as a low-energy effective theory with the cutoff scale $\Lambda$.

The model (\ref{eq:5}) generically admits a 
flat-space ``vacuum'' solution
\[
g_{\mu\nu} = \eta_{\mu\nu};\quad \phi^0=\Lambda^2 t;\quad
\phi^i=\Lambda^2 x^i.
\]
The vacuum possesses rotational symmetry provided the function
$\mathcal{F}$ is invariant under the rotations of the fields $\phi^i$
in the internal space. The Lorentz symmetry is, in general, broken. 
Requiring that the action (\ref{eq:5}) is invariant under the
following symmetry, 
\begin{equation}
\phi^{i}\rightarrow\phi^{i}+\Xi^{i}\left(\phi^{0}\right), 
\label{eq:symmetry}
\end{equation}
where $\Xi^{i}$ are arbitrary functions of $\phi^{0}$, ensures that
perturbations about this solution contain only two propagating
degrees of freedom \cite{Dubovsky:2004sg} which are two polarizations
of a massive graviton with the mass of order $m=\Lambda^2/M_{\rm
  Pl}$, where $M_{\rm Pl}$ is the Planck mass. The model does not 
contradict the most obvious experimental constraints
\cite{Dubovsky:2005dw,Dubovsky:2004ud,Bebronne:2007qh} for graviton
masses as large as $10^{-20}$~eV. 

The ansatz for the static spherically symmetric solution can be
written in the following form \cite{key-5}:
\begin{align}
ds^{2}&=\alpha(r)\text{d}t^{2}-\beta(r)\text{d}r^2
-r^{2}\left(\text{d}\theta^{2}+\sin^{2}
\theta \text{d}\phi^{2} \right), \nonumber \\
\phi^{0}&=\Lambda^{2}\left[t+h(r)\right],\nonumber \\
\phi^{i}&=\phi(r)\frac{\Lambda^2 x^{i}}{r}, \label{eq:10}
\end{align}
where the coordinate transformations $r\rightarrow r'=r'(r)$ and
$t\rightarrow t'=t+\tau(r)$ have been used to eliminate some of the
fields. 

The analytical black hole solutions can be obtained if the function
$\mathcal{F}$ has the form 
\[
\mathcal{F} = \frac{12}{\lambda X}  + 6\left(\frac{2}{\lambda}+1\right)w_1 
- w_1^3 +3 w_1w_2 - 2w_3 +12,
\]
where $\lambda$ is a positive constant and 
\begin{align}
X&= g^{\mu\nu}\partial_\mu \phi^0 \partial_\nu\phi^0/\Lambda^4, \quad
w_n = {\rm Tr} (W^{ij})^n,
\nonumber \\
W^{ij} &= \Lambda^{-4} \partial^\mu \phi^i \partial_\mu \phi^j
- \frac{ \partial^\mu \phi^i \partial_\mu \phi^0   
\partial^\nu \phi^j \partial_\nu \phi^0}{\Lambda^8 X}.
\nonumber
\end{align}
Note that the dependence of $\mathcal{F}$ on the scalar fields through
two combinations $X$ and $W^{ij}$ ensures the symmetry
(\ref{eq:symmetry}). 

With this function $\mathcal{F}$, the black hole solution reads
\begin{align}
\alpha(r)&=1-\frac{2 M G_N}{r}-\frac{Q}{r^{\lambda}},\nonumber\\
\beta(r)&=\frac{1}{\alpha(r)},\nonumber\\
h(r)&=\pm \int \frac{dr}{\alpha}\left[1
-\alpha \left(\frac{Q}{12m^2}\frac{\lambda(\lambda-1)}{r^{\lambda+2}}
+1\right)\right]^{1/2},\nonumber\\
\phi(r)&=r.  \label{eq:16}
\end{align}
Here $M$ and $Q$ are two arbitrary integration constants. We will
assume in what follows that $\lambda>1$. In this case the asymptotics
of the gravitational potential is Newtonian and is given by the parameter
$M$ which determines the ADM mass of the solution. 

\begin{figure}
\begin{minipage}{0.32\linewidth}
\centering
\includegraphics[scale=0.55]{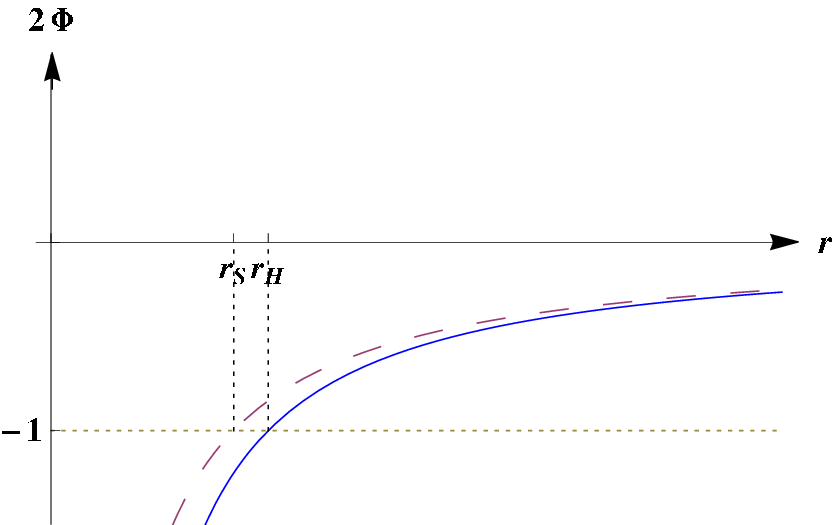}
\end{minipage}%
\begin{minipage}{0.32\linewidth}
\centering
\includegraphics[scale=0.55]{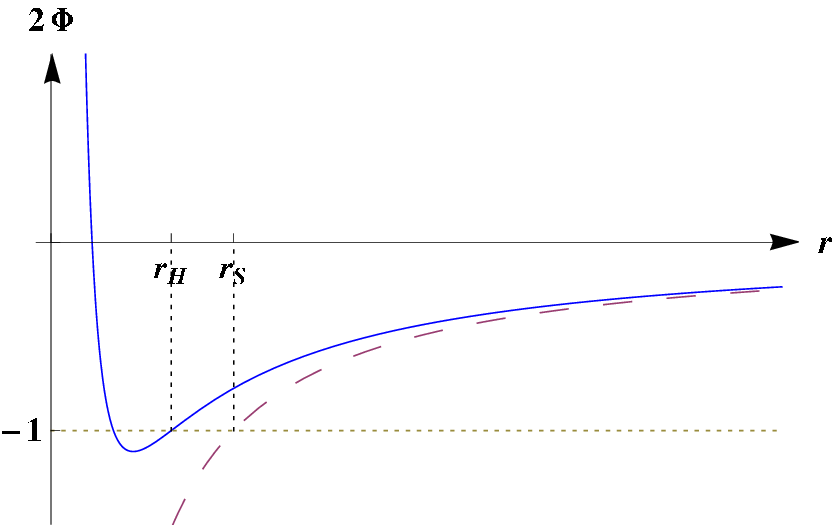}
\end{minipage}%
\begin{minipage}{0.32\linewidth}
\centering
\includegraphics[scale=0.55]{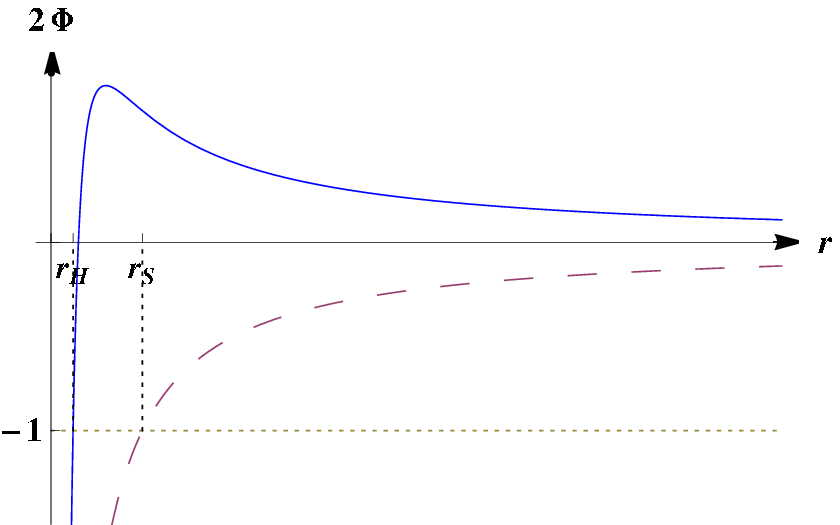}
\end{minipage}%
\caption[Newtonian potential $2\Phi = g_{00}-1$ in the three
     different cases: $M>0$ and $Q>0$, $M>0$ and $Q<0$, $M<0$ and
     $Q>0$ (from left to right). For comparison, the potential of the
     Schwarzschild solution of the mass $|M|$ is shown by the
     dashed line.]{Newtonian potential $2\Phi = g_{00}-1$ in the three
     different cases: $M>0$ and $Q>0$, $M>0$ and $Q<0$, $M<0$ and
     $Q>0$. For comparison, the potential of the
     Schwarzschild solution of the mass $|M|$ is shown by the
     dashed line.} \label{fig:1}
\end{figure}

The solution may possess an horizon which, in this case, is given by
the largest root of the equation 
\begin{equation}
\frac{2 M G_N}{r_H}+\frac{Q}{r_H^{\lambda}}=1
\label{eq:horizon}
\end{equation}
Depending on the signs and relative values of the parameters $M$ and
$Q$, there are three different cases when the horizon exists:
($M>0,Q>0$), $(M>0,Q<0)$ and $(M<0,Q>0)$.The three corresponding
solutions are shown in Fig.~\ref{fig:1} 

In the case $M>0$ and $Q>0$, the black hole has an attractive
gravitational potential at all distances. The attraction is stronger
than that of the usual Schwarzschild black hole of mass $M$; The
horizon size of the modified black hole is also larger (left panel of
Fig.~\ref{fig:1}).

When $M>0$ and $Q<0$, the horizon only exists when the
condition
\begin{equation}
2 M G_N \geq \lambda \left|Q\right|^{1/\lambda} 
\left(\frac{1}{\lambda-1}\right)^{\frac{\lambda-1}{\lambda}} 
\label{eq:17-a}
\end{equation}
is fulfilled. In this case, the Newton's potential is also always
attractive. However, the attraction is weaker than in the
Schwarzschild case, and the horizon size is smaller.

Finally, in the case $M<0$ and $Q>0$ the Newton's potential is
repulsive at large distances and attractive near the horizon. This
case is interesting since it doesn't exist in GR. Although formally
the black hole mass in GR can also be taken negative, this does not
correspond to a physical solution because of the naked singularity.
Another reason to disregard such solutions is the null energy condition
which holds for the matter stress tensor
\cite{key-8,key-9,key-10,key-10.bis}. Neither of the arguments exist in the
modified gravity case. At small distances the repulsion changes to the
attraction, which creates the event horizon hiding the singularity
(right panel of Fig.~\ref{fig:1}). Also, massive gravity doesn't
satisfy the null energy condition, allowing for negative mass states
to be constructed, e.g., as in the ghost condensate model
\cite{key-11}.

\section{The Thermodynamical Properties of the Black Hole Solutions}
In order to analyse the validity of the second law of black hole
thermodynamics we first compute the temperature and entropy of the
modified black holes. We use the approach by Wald
\cite{key-12,key-13,key-14,key-15,key-16} based on the Noether's
charge. As we will show, the entropy is given by the
Bekenstein-Hawking formula, that is equal, in the appropriate units,
to the one quarter of the horizon area.

\subsection{The Black Hole Temperature}

Due to quantum particle creation, black holes emit thermal radiation
\cite{key-0.3}. The temperature of this radiation is determined by the
surface gravity $\kappa$ of the black hole, that is, the acceleration
experienced by a test body at the black hole horizon. The event
horizon of a static spherically symmetric black hole corresponds to a
killing horizon, i.e, a surface to which a killing vector field is
normal (bifurcation surface). The surface gravity $\kappa$ at any
point of a killing horizon $\mathcal{H}$ is defined by
\begin{equation}
\xi^{a}\nabla_{a}\xi^{b}=\kappa \xi^{b}, 
\label{eq:17}
\end{equation}
where $\xi^{a}$ is the killing field normal to $\mathcal{H}$. For a
static spherically symmetric metric of the form
\begin{equation}
ds^{2}=\alpha(r)\text{d}t^2-\frac{1}{\alpha(r)}\text{d}r^2
-r^{2}\left(\text{d}\theta^{2}+\sin^{2}\theta \text{d}\phi\right),
\label{eq:18}
\end{equation}
the surface gravity equals
\begin{equation}
\kappa= \left.\frac{\alpha'(r)}{2}\right|_{r=r_{H}},\label{eq:19}
\end{equation}
where $r_{H}$ is the horizon radius. The zeroth law of
black hole mechanics asserts that $\kappa$ is constant over the event
horizon of a stationary black hole. Although in GR one needs to use
the Einstein's field equations to prove this statement, and thus its
generalizations to other theories of gravity is questionable, the
zeroth law trivially holds for spherically symmetric black holes.

Making use of the explicit solution (\ref{eq:16}) one finds 
\begin{align}
T_H&=\frac{\kappa}{2\pi}=\frac{1}{4\pi r_{H}}
+\frac{1}{4\pi}(\lambda-1)\frac{Q}{r_{H}^{\lambda+1}} 
\nonumber\\
&=\frac{1}{4\pi}\left(\frac{1+(\lambda-1)Q
r_{H}^{-\lambda}}{r_{H}}\right). 
\label{eq:38}
\end{align}
One recovers the temperature of the Schwarzschild black hole in the
limit of zero scalar charge, in which case $T_H=1/4\pi r_H$ as
expected. Moreover, since the temperature of the black hole
corresponds, from a mathematical point of view, to the tangent of the
Newton's potential at the event horizon, it is easy to conclude from
Fig.\ref{fig:1} that the existence of an event horizon implies
$T_{H}\geq 0$.

\begin{figure}
\centering
\includegraphics[scale=0.9]{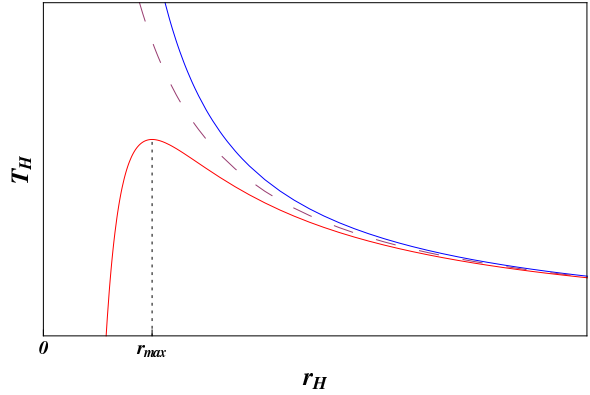}
\caption{The red line corresponds to the temperature of a black hole
  with negative scalar charge. It reaches its maximum at
  $r_{\text{max}}=\left[(\lambda^{2}-1)|Q|\right]^{1/\lambda}$. The
  blue line represents the temperature as a function of the horizon
  radius for positive scalar charge solutions. The dashed line is the
  Schwarzschild case.} \label{fig:2}
\end{figure}

Interestingly, the Hawking temperature behaves differently for
positive and negative scalar charges. For positive $Q$, the
temperature is larger than at $Q=0$ and decreases with $r_H$ as in the
case of a conventional Schwarzschild black hole. On the contrary, at
$Q<0$ the temperature is smaller than in the Schwarzschild
case. Moreover, its dependence on $r_H$ is not monotonic, with a
maximum reached for 
$r_H = r_{\text{max}}=\left[(\lambda^{2}-1)|Q|\right]^{1/\lambda}$ (see
Fig.~\ref{fig:2}). When $r_H>r_\text{max}$, the specific heat is
negative, as in the Schwarzschild case, corresponding to the fact that
the black hole heats up as it radiates energy, while when
$r_H<r_\text{max}$, the specific heat is positive, which means that
the black hole cools down as it radiates, eventually reaching an
equilibrium. 

Another interesting observation is that the temperature reaches
zero at a finite event horizon radius for $Q<0$. This is closely
related to the third law of black hole thermodynamics and will be
discussed in the next section.

\subsection{The Noether's Charge as the Entropy}

As has been proven in Refs.~\cite{key-12,key-13,key-14,key-15,key-16}
by assuming a theory which admits black holes with a bifurcate horizon
and a well-defined mass at infinity, the entropy of the black
hole is given by 
\begin{equation}
S_{H}=2\pi \int_{\mathcal{H}}{\bf Q}[\xi], 
\label{eq:20}
\end{equation}
where ${\bf Q}[\xi]$ is the Noether charge related to the Killing
field $\xi^{a}$ normal to the horizon $\mathcal{H}$. 
For a general theory with matter fields $\psi$ and the Lagrangian
\begin{equation}
\mathcal{L}=\mathcal{L}\left(g_{ab},\mathcal{R}_{abcd},\psi,
\nabla_{a_{1}}\psi,\nabla_{(a_{1},\dots,}\nabla_{a_{j})}\psi\right),
\label{eq:22}
\end{equation}
this quantity can be recast in the following form
\cite{key-13}:
\begin{equation}
S_{H}=-2\pi\int_{\mathcal{H}}d^{D-2}x\sqrt{\sigma}
\frac{\delta \mathcal{L}}{\delta
\mathcal{R}_{abcd}}\hat{\epsilon}_{ab}
\hat{\epsilon}_{cd}, 
\label{eq:21}
\end{equation}
where $\sigma$ is the determinant of the metric on $\mathcal{H}$ and
$\hat{\epsilon}_{ab}$ is the binormal vector to the bifurcation
surface. The black hole entropy is then related to the Noether's
charge of diffeomorphism under the Killing vector field which produces
the event horizon.

Eq.~(\ref{eq:21}) can be directly applied to  the case of massive
gravity with the action \eqref{eq:5}. For metric tensors of the form
\eqref{eq:18}, the relevant Killing vector is $\partial_{t}$, while
the binormal vector $\hat{\epsilon}_{ab}$ has the following components:
$\hat{\epsilon}_{tr}=-\hat{\epsilon}_{rt}=1$ and the other components
vanish. The explicit expression for the entropy is then 
\begin{align}
S_{H}&=-2\pi \oint_{r=r_{H}} \left[ \left(
\frac{\partial \mathcal{L}}{\partial \mathcal{R}_{abcd}}\right)^{(s)}
\hat{\epsilon}_{ab}\hat{\epsilon}_{cd}r^{2}\sin \theta\right]
\text{d}\theta \text{d}\phi \nonumber \\
&=-8\pi \oint_{r=r_{H}} \left[\left(\frac{\partial 
\mathcal{L}}{\partial \mathcal{R}_{rtrt}} \right)^{(s)}r^{2}
\sin\theta\right]\text{d}\theta \text{d}\phi, 
\label{eq:23}
\end{align}
where the factor 4 is a consequence of the antisymmetry property of
the Riemann tensor and the binormal vectors. The index (s) is to
emphasize that the functional is evaluated on the solution
\eqref{eq:16}. Substituting the Lagrangian \eqref{eq:5} and making use
of the fact that the function $\mathcal{F}$ does not dependent on the
Riemann tensor (the metric and the Riemann tensor are treated as
independent variables in eq~(\ref{eq:23})), we arrive at the
Hawking-Bekenstein formula :
\begin{align}
S_{H} &= -8\pi \oint_{r=r_{H}} \left[\frac{1}{32\pi}
\left(g^{tt}g^{rr}-g^{tr}g^{tr}\right)\right]r^{2}
\sin \theta \text{d}\theta \text{d}\phi, 
\nonumber \\
&=\frac{1}{4}\left[\int_{0}^{\pi} \int_{0}^{2\pi}r^{2}
\sin \theta \text{d}\theta \text{d}\phi\right]_{r=r_{H}} =\frac{A_{H}}{4}, 
\label{eq:24}
\end{align}
where $A_H$ is the area of the black hole horizon in the Planck units
$G_{N}=1$.

As it has been shown in Ref.\cite{key-12}, this expression for the
entropy is automatically consistent with the first law of
thermodynamics. It can also be checked directly by making use of the
expressions (\ref{eq:24}) and (\ref{eq:horizon}) at $dQ=0$.

\subsection{The Second and Third Laws of Thermodynamics}

Let's first examine the generalized second law of thermodynamics. To
this end consider the change of entropy in the process of coalescence
of two black holes characterized by the masses $M_{1}$ and $m_{2}$ and
scalar charges $Q_{1}$ and $q_{2}$, respectively. Let both black holes
be large enough so that the Hawking radiation can be neglected and
their horizon radii are larger than the inverse cutoff scale, in which
case the solutions are within the region of validity of the effective
theory.

Before the coalescence, when the interaction of the two black holes
can be neglected, their entropy is simply the sum of the entropies of
the two isolated black holes as given by eqs.~(\ref{eq:24}) and
(\ref{eq:horizon}). In order to check the second law of thermodynamics
we need to determine the entropy of the final state. Let us argue that
in the limit when one of the black holes is much larger than the
other, and the scalar charges of the black holes are not
parametrically larger than their masses,
\begin{equation}
|M_1|\gg |m_2|,\quad |Q_1|^{1/\lambda}\lesssim |M_1|,
\quad |q_2|^{1/\lambda}\lesssim |m_2|,
\label{eq:limit}
\end{equation}
the result of the coalescence is a black hole with the mass $M_1+m_2$
and the scalar charge $Q_1+q_2$. Then the entropy of the final state
is given by eqs.~(\ref{eq:24}) and (\ref{eq:horizon}), and the net
change of entropy can be easily calculated.

The argument consists of two parts. First, let us show that in the
above limit the total energy of gravity waves emitted in the process
of coalescence is parametrically small as compared to the mass of the
smallest black hole. Assume that the characteristic time scale of the
coalescence is much smaller than the inverse graviton mass (this can
always be arranged in view of the tiny value of the latter). Then the
gravitational waves can be considered massless.  The metric
perturbation $h_{ij}$ in the case of the quadrupole radiation is
estimated as follows:
\begin{equation}
h_{ij}\sim \frac{\ddot{Q}_{ij}}{M_{pl}^{2}r}, 
\label{eq:30}
\end{equation}
where $Q_{ij}$ is the second time derivative of the quadrupole moment
and $r$ is the characteristic size of the system. The energy density
in the gravitational waves is, therefore, of order
\begin{equation}
\rho \sim \omega^2 M_{pl}^{2} h_{ij}^2
\sim \omega^2 \frac{(\ddot{Q}_{ij})^2}{M_{pl}^2r^2},
\label{eq:31}
\end{equation}
where $\omega$ is the characteristic frequency of the emitted waves
as seen by an asymptotic observer. Setting the size of the system to
$r\sim R_H$ and the frequency to $\omega\sim 1/R_H$, $R_H$ being the
horizon size of the large black hole, we have 
\[
\ddot{Q}\sim\omega^{2}Q\sim\omega^{2}m_{2}R_{H}^{2}
\]
and, therefore, the total energy emitted over the time period $\sim R_H$ is 
\begin{equation}
E_{rad}\sim m_{2}\left(\frac{m_{2}}{M_{1}}\right) 
\sim  m_{2}\left(\frac{r_{H}}{R_{H}}\right),
\label{eq:Erad}
\end{equation}
where $r_H$ is the horizon size of the small black hole and we have
assumed that the presence of scalar charges does not change completely
the horizon sizes of the two original black holes. We see from
eq.~(\ref{eq:Erad}) that in the limit $m_2\ll M_1$ the gravitational
radiation during the coalescence of the two black holes can be
neglected. 

We now turn to determining the mass and the scalar charge of the
resulting black hole. This can be done by considering the asymptotic
gravitational potential created by the coalescing black holes and
extracting the coefficients in front of the $1/r$ and $1/r^\lambda$
terms. These coefficients determine the mass and the scalar charge of
the resulting black hole, respectively. 

To clarify the logic, consider first the coalescence of the two black
holes of masses $M_1$ and $m_2$ in the conventional GR. In this case
the asymptotic gravitational potential of each black hole is
determined by its mass and satisfies the linear equation. Thus, before the
coalescence and to the leading order in $1/r$, the gravitational
potential of the two black holes is given by the sum of the potentials
of the individual black holes. The coefficient of $1/r$ is independent
of the distance between the black holes (as long as this distance is
much smaller than $r$) and is given by the sum of the black hole
masses. Since after the coalescence this coefficient is determined by
the mass $M$ of the resulting black hole and there is no substantial
emission of the gravitational waves, we conclude that $M=M_1+m_2$.

In the case of the modified black holes the situation is more
complicated because the corresponding solution is non-linear at all
distances. This introduces a correction to the simple addition
of masses and charges, 
\[
M = M_1+m_2+\delta m; \quad Q=Q_1+q_2+\delta q.
\]
However, one can choose the parameters $M_1$, $m_2$, $Q_1$ and $q_2$
in such a way that these corrections are negligible. Making use of an
exactly treatable spherically symmetric case to estimate $\delta q$
and $\delta m$ (see Appendix \ref{sec:append}), one may argue that
this can be achieved by requiring that
\begin{equation}
\left(\frac{r_H}{R_H}\right)^{1+1/\lambda}
\gg \left(\frac{\mu}{M_{pl}}\right)^{3/\lambda}.
\label{eq:32}
\end{equation}
The last condition can always be satisfied for a sufficiently small
value of the graviton mass $\mu$ without leaving the region of
validity of the effective theory. In this case, the effects of
non-linearity on the asymptotic potential can be neglected and the
arguments presented above imply that the resulting black hole has, to
a good accuracy, the mass $M=M_1+m_2$ and the scalar charge
$Q=Q_1+q_2$.

Having determined the mass and the scalar charge of the final black
hole, we can now calculate the change of the entropy in the process of
coalescence,
\begin{align}
\triangle S_{bh}&=S_{\text{final}}-S_{\text{initial}}=\pi 
\left(\mathcal{R}_{H}^{2}-R_{H}^{2}-r_{H}^{2}\right) 
\nonumber\\
&\approx \pi \left(\mathcal{R}_{H}^{2}-R_{H}^{2}\right), 
\label{eq:33}
\end{align}
where $\mathcal{R}_{H}$ is the radius of the event horizon of the
final black hole. 

In the limit (\ref{eq:limit}) the entropy change can be calculated
expanding in powers of $m_2/M_1$. To the leading order, the answer is
linear in the mass of the small black hole, 
\begin{equation}
\triangle S_{bh}\simeq \frac{4\pi
R_{H} m_2 }{1+(\lambda-1)Q_{1}
R_{H}^{-\lambda}} = \frac{m_{2}}{T_{H}},
\label{eq:37}
\end{equation}
where $T_H$ is the temperature of the large black hole.  Thus, the
entropy may increase or decrease, depending on the sign of the mass of
the small black hole. For negative masses, the generalized second law
of thermodynamics is violated.

The decrease of the entropy when negative mass black holes are
involved can be checked explicitely in the particular case of
$\lambda=2$. In this case eq.~(\ref{eq:horizon}) for the
horizon in terms of the mass and the scalar charge can be solved
exactly,
\begin{equation}
R_{H}=M+\sqrt{M^2 +Q}. 
\label{eq:40}
\end{equation}
The existence of the horizon requires $M^2+Q\geq 0$, which is always
the case for a positive scalar charge. For negative $Q$, the condition
\eqref{eq:17-a} has to be imposed. 

The change of the black hole entropy can be computed
explicitly giving 
\begin{align}
\triangle S_{bh}=&2\pi \biggl[ (M_{1}+m_2) 
\sqrt{(M_1+m_2)^2+Q_{1}+q_{2}} 
\nonumber \\
& + 2M_{1}m_{2}-M_1\sqrt{M_{1}^2+Q_{1}}  
-m_2\sqrt{m_{2}^{2}+q_{2}} \biggr].
\label{eq:41}
\end{align}
One can see from this expression that the entropy always increases
when both black holes have positive masses. However, when at least one
of the masses is negative and $M_{1}\gg |m_{2}|$, one arrives at the
same conclusion as before: there is a decrease of entropy and the
generalized second law of thermodynamics is violated.
 
Consider now the third law of thermodynamics. In application to black
holes, it states that a black hole with a non-vanishing temperature
cannot reach zero temperature in a finite sequence of operations
\cite{key-24}. In our case, the zero temperature corresponds to an
``extremal'' situation when the inequality in eq.~(\ref{eq:17-a})
becomes an equality. Indeed, the black hole temperature can be
expressed in terms of the surface gravity which, according to
eq.~(\ref{eq:19}), is proportional to the slope of $\alpha(r)$ at the
horizon. It is then clear from Fig.~\ref{fig:1} that the zero
temperature can only be reached in the case $M>0$, $Q<0$ (represented
on the middle panel) when the two roots of eq.~(\ref{eq:horizon})
coincide. Thus, an extremal black hole satisfies the condition
\begin{equation}
2M=\gamma |Q|^{\beta}
\label{eq:zeroT}
\end{equation}
with $\beta=1/\lambda$ and
\[
 \gamma=\lambda\left(\frac{1}{\lambda-1}\right)^{
\frac{\lambda-1}{\lambda}}>0.
\]
Note that both $\gamma$ and $\beta$ are positive real numbers. For
simplicity, we will concentrate on the case $\lambda=2$ from now
on. In this case, $\beta=1/2$ and $\gamma=2$. The generalization to
other cases is straightforward.

Consider again the coalescence of two black holes. Let the large black
hole be nearly extremal, so that $Q_1<0$ and 
\[
M_{1}=(1+\varepsilon)|Q_{1}|^{1/2},
\]
where $\varepsilon$ is a small positive number ($\varepsilon=0$
corresponds to the extremal black hole). If the small black hole has a
positive mass and a scalar charge satisfying the condition
(\ref{eq:17-a}), it is easy to see that the black hole resulting from
coalescence will also satisfy the condition (\ref{eq:17-a}) and thus
will have a non-zero temperature. Indeed, the final state is
characterized by the inequality
\[
(M_{1}+m_{2})^2+Q_{1}+q_{2}=(M_{1}^2+Q_{1})+(m_{2}^2+q_{2})+2m_{2}M_{1}>0.
\]
On the contrary, if the small black hole has a negative mass, one can
always choose its charge such that it converts the large black hole
into an extremal one.  For example, the mass of the small black hole
can be taken $m_{2}=-2\varepsilon |Q_{1}|^{1/2}$.  The condition $m_2
\ll M_1$ is still fulfilled for small enough values of $\varepsilon$.
The scalar charge of the second black hole can be chosen as
$q_{2}=\delta |Q_{1}|$ with $\delta$ a small positive number. Then,
the final state is a black hole with mass
$\mathcal{M}=(1-\varepsilon)|Q_{1}|^{1/2}$ and a scalar charge
$|\mathcal{Q}|=(1-\delta)|Q_{1}|$, which means that
$\mathcal{M}=|\mathcal{Q}|^{1/2}$ if we choose
$(1-\varepsilon)=(1-\delta)^{1/2}$.  A non-extremal black hole has
turned into an extremal one in a one-step process.  Thus, when
negative mass black holes are present, the third law of thermodynamics
is not fulfilled.

The third law of black hole thermodynamics is closely related to the
cosmic censorship conjecture \cite{key-20}, which states that all
singularities resulting from a gravitational collapse are always
hidden by the horizon. With an obvious modification, the above process
that leads to an extreme black hole can be used to create a naked
singularity. This possibility is again related to the existence of the
negative-mass black holes and, ultimately, to the violation of the
null energy condition in massive gravity.

\section{Discussion}

In this paper we have addressed the validity of the generalized laws
of thermodynamics in massive gravity models. In these models the black
hole solutions are modified by the presence of the scalar
``hair''. The analog of the Schwarzschild black hole -- the
spherically symmetric solution -- depends on two parameters, the mass
and the ``scalar charge'' characterizing the hair strength. The
presence of two free parameters makes the asymptotics of the
gravitational field of the black hole essentially independent of its
behavior near the horizon, allowing for negative-mass solutions
without naked singularity.

Making use of the exact black hole solutions, we have constructed
explicit examples of the black hole mergers which violate the
generalized second and third laws of thermodynamics. The existence of
such processes is in accord with the general theorems of the black
hole thermodynamics which require the weak energy condition to
hold. The latter condition, together with the null energy condition,
is violated in massive gravity models.  Indeed, the examples we have
constructed involve the negative-mass black holes. Note that the
violation of the null energy condition in massive gravity is related
to the presence of the superluminal modes \cite{Dubovsky:2005xd} which
are likely to be responsible for the existence of the black hole hair
\cite{Dubovsky:2007zi}. 

Although the negative mass black holes are classical solutions of
massive gravity which exist on the same footing as the conventional
positive-mass black holes, this does not guarantee that they can be
created from positive-mass states in the course of the evolution. If
this were the case, the violation of the second law would imply that
one may devise a process involving black holes which would
convert heat into work. Moreover, the violation of the third law would
mean that the cosmic censorship conjecture is not true, so that one
can create naked singularities starting from singularity-free states.
 
It is not inconceivable that the creation of the negative-mass states
can be forbidden dynamically. In fact, it has been conjectured in the context of the ghost
condensate model that an average null energy condition (ANEC) may 
prevent the entropy of the black hole from decreasing, in a
coarse-grained sense \cite{key-4}. This means that the second law
of thermodynamics may be violated locally, while in average (during
long time periods) it holds. The ANEC can
be traced back to the presence of the conserved Noether's charge
related to the shift symmetry of the ghost condensate field. 
In our case, a similar internal shift symmetry
is also present, so one may wonder whether this symmetry implies 
ANEC which may prevent the breakdown of thermodynamical laws.

The situation, however, is not exactly the same in massive gravity models.
First, massive gravity possesses four scalar fields $\phi^0$
and $\phi^i$, the former being an analog of the ghost condensate field. Even though
these fields decouple from the ordinary matter, they are coupled to each other, so 
the scalar sector is more complicated than in the ghost condensate model. For this 
reason we have not been able to directly generalize the argument based on ANEC, 
except for linear perturbations above the Minkowski background.
This, however, is not sufficient to argue against the formation of the negative-mass black
holes, so the question remains open.

Moreover, the original argument \cite{key-4} is partially based on the fact 
that a negative Noether's charge is strongly disfavored since it is leading
to UV instabilities. However, the massive gravity is free from UV 
instabilities, at least for backgrounds which can be approximated
as flat in the UV limit (see \cite{Dubovsky:2005dw} for details). 
This implies that part of the argument in \cite{key-4} doesn't go
through in our case. 

The final observation is that ANEC cannot protect the third law of thermodynamics and the cosmic
censorship conjecture. Indeed, unlike the entropy which may decrease locally but still increase in
average, the naked singularity, once created, cannot be undone no matter what happens in other
places. The breakdown of the cosmic censorship conjecture may therefore occur even if ANEC holds. 
The creation of naked singularities may signal the problem of the model, unless they are somehow 
regularized by an appropriate UV-completion.

\acknowledgments

We are grateful to M. Bebronne, S. Dubovsky and S. Sibiryakov for
valuable discussions and comments on the manuscript.  This work is
supported in part by the IISN, Belgian Science Policy (under contract
IAP V/27).  The work of P.T. is supported in part by the ``Action de
Recherche Concert\'ees'' (ARC), project "Beyond Einstein: fundamental
aspects of gravitational interactions".

\appendix
\section{Estimate of charge and mass correction} 
\label{sec:append}
Since the equations of massive gravity do not linearize asymptotically
because of the slow decay of the field $\phi^0$ \cite{key-5}, the state of
two close black holes with parameters $M_1$, $Q_1$ and $m_2$, $q_2$
has asymptotic mass and scalar charge (i.e., the coefficients in
front of $1/r$ and $1/r^\lambda$ in the Newtonian potential) equal to 
\[
M=M_1+m_2+\delta m, \qquad Q=Q_1+q_2+\delta q. 
\]
Our goal here is to estimate the corrections $\delta m$ and $\delta q$
and find the parameters for which these corrections can be neglected.

For the estimate we make use of the fact that for a spherically
symmetric matter distribution the equations can be solved exactly. We
model the large black hole by a sphere of constant density of the
total mass $M_1$ and radius close to the horizon size $R\sim 2G_NM_1$,
and a small black hole by a spherical shell of the same density and
the mass $m_2$, covering the sphere $R$. Even though the geometry
is different, we expect that our model configuration reproduces
correctly the parametrical dependence of $\delta m$ and $\delta q$ on
$R$, $M_1$ and $m_2$. 

In a spherically symmetric case it is straightforward to obtain the
scalar charge as of a sphere of a constant density as a function of
the mass $M_1$ and radius $R$ \cite{Comelli:2010bj}. Matching
boundary conditions, we obtain the following relation for the scalar
charge
\begin{equation}
 Q_1=\mathcal{C}(\lambda)M_1G_{N}\mu^2 R^{1+\lambda},
\label{eq:Q1=}
\end{equation}
where $\mathcal{C}(\lambda)$ is a constant depending on $\lambda$
which is of order one for $\lambda$ varying between one and two, and
$\mu$ is the graviton mass.

Making use of eq.~(\ref{eq:Q1=}), one may compute the change of the scalar
charge when a thin shell around the constant density sphere is added. 
The result is 
\[
\delta q \sim m_2 G_{N}\mu^2 R^{1+\lambda}.
\]
This estimate can be rewritten in terms of the parameters of the two
black holes as follows: 
\[
\delta q \sim \mu^2 r_H ~ R_H^{\lambda+1}.
\]
Requiring $\delta q\ll q_2$ we arrive at the condition
(\ref{eq:32}) for the worst case scenario, i.e. when $m_2\sim\Lambda$.
If the condition (\ref{eq:32}) is satisfied, it automatically implies that $\delta  m \ll
m_2$.

\end{document}